\documentclass[12pt]{article}
\usepackage{amsfonts}
\usepackage{psfrag}

\usepackage{amsmath,amssymb}
\usepackage[dvips]{graphicx}

\begin{document}

\begin{center}
\textbf{{\large Perturbative Quantum Gravity Coupled to Particles in $(1+1)$-Dimensions}}\\[0pt]

\vspace{48pt} R.B. Mann$^{(1)}$, M. B. Young$^{(1,2)}$ \vspace{12pt}

\textit{$^1$ Department of Physics \& Astronomy, University of Waterloo, \\[0pt]
Waterloo, Ontario, N2L 3G1, Canada }

\smallskip \textit{$^2$ Department of Mathematics and Statistics, Queen's
University at Kingston, \\[0pt]
Kingston, Ontario, K7L 3N6, Canada } \bigskip

mann@avatar.uwaterloo.ca, 3my@qlink.queensu.ca

\vspace{24pt}
\end{center}
\begin{abstract}
\addtolength{\baselineskip}{1.2mm} \addtolength{\baselineskip}{1.2mm}
We consider the problem of $(1+1)$-dimensional quantum gravity coupled to particles.
Working with the canonically reduced Hamiltonian, we obtain its
post-Newtonian expansion for two identical particles.  We quantize the $(1+1)$-dimensional
Newtonian system first, after which explicit
energy corrections to second order in $c^{-1}$ are obtained. We compute the perturbed
wavefunctions and show that the particles are bound less tightly together than in the Newtonian case.
\end{abstract}

\addtolength{\baselineskip}{1.5mm} \thispagestyle{empty}

\vspace{48pt}

\setcounter{footnote}{0}

\section{Introduction}

It is safe to say that we are still a considerable distance from a satisfactory
theory of quantum gravity.  Pursuit of this goal is taking place on many fronts,
one of which involves working in a two dimensional setting. This furnishes a dynamically simplified framework
for developing methods that might be generalizable to higher dimensional quantum gravity,
as well as yielding potential insight into quantization of a restricted number of degrees of
freedom of such theories.

Although
the Einstein vacuum field equations in two spacetime dimensions are trivial insofar
as they are satisfied by any metric, it is nevertheless possible to construct
dynamically interesting theories in this setting.  This can be done by coupling a scalar field
(called the dilaton) into the theory and including suitable matter sources, which
typically include the dilaton itself.  This has the unfortunate drawback that
the theory under consideration typically does not resemble general relativity in
its lower-dimensional setting, although some two-dimensional theories are equivalent to a reduction of higher-dimensional Einstein gravity to spherical symmetry \cite{2dreview}.  However in their own two-dimensional context, such theories
in general do not have a sensible Newtonian limit and/or post-Newtonian expansion
\cite{jchan}.

However it has been shown that one can take the $d\to 2$ limit of general relativity
by suitably rescaling Newton's constant $G$ \cite{Srossd2lim}. The resultant theory
is one in which the Ricci scalar $R$ is set equal to
the trace of the stress energy of prescribed matter fields and sources. In this sense
the theory models (3+1)-dimensional general relativity in a manner not captured by
other dilaton theories of gravity: the evolution of space-time curvature is
governed by the matter distribution, which in turn is governed by the
dynamics of space-time \cite{r3}. Referred to as $R=T$\ theory,  when the stress energy
is that of a cosmological constant, it reduces to Jackiw-Teitelboim theory
\cite{JT}.

An important feature of the $R=T$ theory is that
it has a consistent nonrelativistic ($c\rightarrow \infty $) limit \cite{r3} that yields (1+1)-dimensional
Newtonian gravity. Its quantization  is therefore of considerable potential interest insofar as its (quantum) Newtonian
limit can indirectly be empirically tested.  This is because the effective theory of gravity near the earth's surface is
(1+1)-dimensional Newtonian gravity, whose potential increases linearly with separation away from the surface.
From early experimental tests indicating that spin-1/2 particles undergo gravity-induced phase shifts
\cite{COW} to more recent
work demonstrating  that neutrons experience a quantization of energy levels in
this potential \cite{neutquant}, 2D Newtonian quantum gravity affords an important empirical window into semiclassical quantum gravity.  In attempting to understand (at least qualitatively) how relativistic effects could modify this behaviour the $R=T$ theory can therefore play an important role.

In this paper we take the first steps toward quantization of the $R=T$ theory.  We proceed by taking as
a matter source a  minimally coupled pair of point particles, forming a 2-body relativistic self-gravitating system.
Nonrelativistic (1+1)-dimensional self-gravitating systems (OGS) of N particles have been very important in
the fields of astrophysics and cosmology for over 30 years \cite{Rybicki}, since they not only
furnish prototypes for studying gravity in
higher dimensions, but also provide effective descriptions of physical systems whose dynamics are closely
approximated by the one dimensional system.  The OGS phase space is known to exhibit solutions
corresponding to very long-lived core-halo
configurations, reminiscent of structures observed in globular clusters \cite{yawn}. The OGS also
approximates collisions of flat parallel domain walls moving in directions orthogonal to their
surfaces as well as the dynamics of stars in a direction perpendicular to the plane of
a highly flattened galaxy.  Relativistic self-gravitating systems have only been more recently studied,
primarily in the 2-body \cite{mann1,mannrobprl} and 3-body   \cite{3bdshort,burnell,Justin,koop} cases.

In the 2-body case a broad class of exact solutions have been obtained, with extensions that include
electromagnetism and cosmological constant \cite{2bdchglo}. The Hamiltonian is explicitly known
as a function of the canonical position and momentum variables. As such its quantization would
yield a version of quantum gravity coupled to matter -- perhaps  the simplest non-trivial quantum gravity system -- that affords direct comparison to a classical counterpart.  However the nonlinearities of the Hamiltonian yield
significant challenges, eg. operator-ordering problems.
Here we proceed by quantizing the post-Newtonian expansion of $R=T$ theory. The post-Newtonian
expansion for the N-body system has been previously carried out \cite{mann2}. We consider this
expansion for $N=2$, regarding the post-Newtonian terms as
perturbative corrections to (1+1) Newtonian gravity.  We find that relativistic corrections tend to lower
the ground state energy but raise the energy of all excited states.  Furthermore, such corrections tend
to spread the wavefunction, indicating a less tightly bound system than in the nonrelativistic case.

The outline of our paper is as follows. We begin with a short review of the relativistic $N$-body system, outlining
briefly how the exact Hamiltonian for the relativistic 2-body system is obtained.  We then discuss the post-Newtonian
expansion of this Hamiltonian.  In section 3 we review the quantization of the non-relativistic system and then
in section 4 obtain the post-Newtonian corrections to this system.  In section 5 we consider the limits of
validity of this approximation, and in section 6 finish off with some concluding remarks. An appendix contains
some technical details associated with the perturbative calculations.

\section{$R=T$ Gravity and its Post-Newtonian
Approximation}

We first provide a quick review of the canonical reduction of $(1+1)$-dimensional dilaton gravity.  For the gravitational field coupled with $N$ point particles, the action is \cite{mann2}
\begin{equation}
\begin{array}{rl}
\displaystyle I=&\displaystyle\int dx^2\;\left[\frac{1}{2\kappa}\sqrt{-g}\left(\Psi R+\frac{1}{2}g^{\mu\nu}\nabla_{\mu}\Psi\nabla_{\nu}\Psi\right)\right.\\ & \displaystyle\left.-\sum_{i=1}^N m_i\int d\tau_i\; \left(-g_{\mu\nu}(x)\frac{dz_i^{\mu}}{d\tau_i}\frac{dz_i^{\nu}}{d\tau_i}\right)^{\frac{1}{2}}\delta^2\left(x-z_i(\tau_i)\right)\right].
\end{array}  \label{action}
\end{equation}
Here $g_{\mu\nu}$ is the metric with $g:=\hbox{det}g_{\mu\nu}$, $\Psi$ is the dilaton field, $R$ is the Ricci scalar, $\tau_i$ is the proper time of the $i^{th}$ particle, and $\kappa:=\frac{8\pi G}{c^4}$ is the coupling constant, where $G$ is the gravitational constant in two spacetime dimensions and $c$ is the speed of light.  $\nabla_{\mu}$ is the covariant derivative naturally associated with the metric $g_{\mu\nu}$, $\nabla_{\lambda}g_{\mu\nu}=0$.

From the action (\ref{action}), the field equations for the $R=T$ theory are derived. Upon canonical reduction of the
action and field equations \cite{mann2}, the degrees of freedom of the system are $z_i$ and $p_i$, the position and canonical momentum of the $i^{th}$ particle, respectively. The Hamiltonian density $\mathcal{H}$ is given by
\[
\mathcal{H}=-\frac{1}{\kappa}\Delta\Psi
\]
where $\Delta=\frac{\partial^2}{\partial x^2}$ is the spatial Lapalace operator,
and where $\Psi$ is to be viewed as a function of $z_i$ and $p_i$, determined from the constraint conditions \cite{mann2}
\begin{equation}
\Delta \Psi-\frac{1}{4}\left(\frac{\partial\Psi}{\partial x}\right)^2+\kappa^2\pi^2+\kappa\sum_i\sqrt{p_i^2+m_i^2}\delta(x-z_i)=0 \label{const1}
\end{equation}
and
\begin{equation}
2\frac{\partial\pi}{\partial x}+\sum_ip_i\delta(x-z_i)=0  \label{const2}
\end{equation}
with $\pi$ the conjugate momentum to $g_{11}$.
From this, the reduced Hamiltonian of the system is then
\[
H=\int dx\;\mathcal{H}.
\]

Once the constraint equations (\ref{const1}) and (\ref{const2}) are solved, $H$ can then be written in terms of $z_i$ and $p_i$.  Since the centre of interia momentum is conserved and its conjugate variable does not appear in the
Hamiltonian, the system has only two degrees of freedom: the separation $r:=z_1-z_2$ and the canonical momentum $p$ of the particles. Solving equations (\ref{const1}) and (\ref{const2}) for two identical particles of mass $m$ in the centre of inertia frame gives the Hamiltonian implicilty in terms of these variables
\begin{equation}
\left(H-\sqrt{p^2+m^2}-\epsilon\tilde{p}\right)^2=\left(\sqrt{p^2+m^2}-\epsilon\tilde{p}\right)^2\hbox{exp}\left(\frac{\kappa}{4}\left(H-2\epsilon\tilde{p}\right)|r|\right)  \label{impHam}
\end{equation}
where $\tilde{p}:=\hbox{sgn}(r)p$ and $\epsilon$ is a constant satisfying $\epsilon^2=1$  so as to make  time reversal invariance evident.

\begin{figure}[t]
\psfrag{x}{$x$}
\psfrag{W(x)}{$W(x)$}
\includegraphics[scale=0.250]{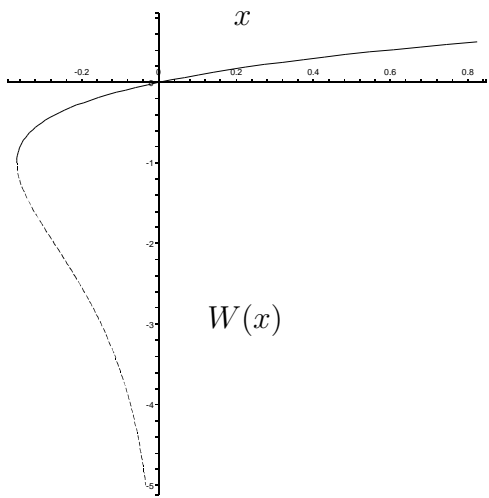}
\caption{The figure shows a plot of both real branches of the Lambert $W$
function, $W_0$ (solid) and $W_{-1}$ (dot). }
\label{lambert}
\end{figure}

Eq. (\ref{impHam}) can be solved explicitly for $H$ giving \cite{mann1}
\begin{equation}
H=\sqrt{p^{2}c^{2}+m^{2}c^{4}}+\hbox{sgn}(r)pc-\frac{c^{4}}{\pi G|r|}W\left(
-se^{s}\right)   \label{exactHam}
\end{equation}%
where
\begin{equation}
s=\frac{\pi G}{c^{4}}\left( |r|\sqrt{p^{2}c^{2}+m^{2}c^{4}}-rpc\right).
\end{equation}%
Note that we have used the explicit form of $\kappa$ and now show the explicit dependence on $c$.
Here, $W(x)$ is the Lambert $W$ function \cite{corle}, defined as the solution $W=y(x)$
to the equation $ye^{y}=x$, which has real solutions for $x\geq -\frac{1}{e}$. $W$ is in general a complex, multivalued function. For $x\in\left[-\frac{1}{e},0\right)$, $W$ has two real branches.  The first, denoted by $W_{-1}$, satisfies $W(x)\leq-1$. The latter, which is defined for $x\in\left[-\frac{1}{e},0 \right)$, satisfies $-1\leq W(x)$, is denoted by $W_0$ and is called the principal branch.  See Figure \ref{lambert} for a plot of both branches.  We will always be dealing with the principal branch $W_{0}$, and will simply write for this $W$.   As suggested in Figure \ref{lambert}, we continue $W_0$ to $x\in\left[-\frac{1}{e},\infty\right)$.

The Hamiltonian (\ref{exactHam}) is exact to all orders in the
gravitational coupling constant $\kappa =\frac{8\pi G}{c^{4}}$, meaning it
accurately describes the system in both the strong and weak field regimes.
Expanding it in powers of $c^{-1}$ yields the post-Newtonian
approximation\footnote{%
As is common in expansions of relativistic expressions, there is a term
proportional to $mc^2$ that represents the rest energy of the system. This
term can be discarded without changing the physics, and hence we will do so
from now on.}
\begin{equation}
H=2mc^2+\frac{p^2}{m} +2\pi Gm^2|r|-\frac{4\pi Gm}{c}pr-\frac{p^4}{4m^3c^2}+
\frac{4\pi^2 G^2m^3}{c^2}r^2+\frac{4\pi G}{c^2}|r|p^2+\mathcal{O}\left(\frac{%
1}{c^3}\right). \label{postNewt}
\end{equation}
This gives an approximation of Eq. (\ref{exactHam}) in the case of slow
motion in a weak gravitational field. The Hamiltonian (\ref{postNewt})
contains a term proportional to $c^{-1}$, which can be removed by making the
coordinate transformation \cite{mann2}
\begin{equation*}
r\mapsto \tilde{r}
\end{equation*}
\begin{equation*}
pc\mapsto \tilde{p}c+2\pi Gm^2\tilde{r}.
\end{equation*}
The Hamiltonian (\ref{postNewt}) then becomes
\begin{equation}
\tilde{H}=\frac{\tilde{p}^2}{m} +2\pi Gm^2|\tilde{r}|-\frac{\tilde{p}^4}{%
4m^3c^2}+\frac{4\pi G}{c^2}|\tilde{r}|\tilde{p}^2 +\mathcal{O}\left(\frac{1}{%
c^3}\right).  \label{hamil}
\end{equation}
The above transformation is easily seen to be canonical by the anti-symmetry
and linearity of the Poisson brackets, and hence $\displaystyle\tilde{H}%
\left(\tilde{r},\tilde{p}\right)$ plays the r\^{o}le of $\displaystyle %
H\left(r,p\right)$ in $\left(\tilde{r},\tilde{p}\right)$ space. \smallskip

\indent In $(3+1)$-dimensional gravity, terms of odd power in $c^{-1}$ can
be associated with gravitational radiation \cite{weinb}. Because there
exists no graviton degree of freedom in the $(1+1)$-dimensional case,
gravitational radiation does not exist, and inclusion of such terms may be
misleading. The transformed Hamiltonian $\tilde{H}$ contains relativistic
terms proportional only to $c^{-2}$, more naturally analogous
to the $(3+1)$-dimensional case. From now on the tildes will be dropped for
clarity with the understanding that only $\left(\tilde{r},\tilde{p}\right)$ coordinates will be used.
\newline
\indent If we define $H^0:=\displaystyle\lim_{c\rightarrow\infty}H$, we see that $H^0$ is the appropriate Hamiltonian to describe two identical
particles in $(1+1)$-dimensional Newtonian gravity. We then have $%
H=H^0+H^{\prime}$ where
\begin{equation}
H^0=\frac{p^2}{m}+2\pi Gm^2|r|  \label{newtHam}
\end{equation}
and
\begin{equation}
H^{\prime}=-\frac{p^4}{4m^3c^2}+\frac{4\pi G}{c^2}|r|p^2.  \label{pertHam}
\end{equation}
In the slow moving limit, for small $r$ and $p$ we see $H\approx H^0$, so
that the relativistic effects are small compared to the Newtonian effects.
This leads to the natural identification of $H^{\prime}$ as a relativistic
perturbation to the Newtonian system.

Note that the Hamiltonian (\ref{exactHam}) is an exact result that fully encapsulates all of
the degrees of freedom of the $R=T$ theory coupled to two point particles.  As such, quantization of
this Hamiltonian is equivalent to a full quantization of the $R=T$ theory coupled to two point particles
in a linear (as opposed to circular) topology. We proceed in the next section to quantize its non-relativistic
counterpart (\ref{newtHam}).

\section{Quantization of  $(1+1)$-Dimensional Newtonian Gravity}

We would first like to gain an understanding of the two particles in the
unperturbed system, described by the Hamiltonian (\ref{newtHam}). Under the
canonical operator identifications in configuration space,
\[
r\mapsto r,\;\;\;p\mapsto -i\hbar\partial_r
\]
this becomes the Hamiltonian operator
\begin{equation}
\hat{H}^0=-\frac{\hbar^2}{m}\partial_r^2+2\pi Gm^2|r|.
\end{equation}
The Schr\"{o}dinger equation then reads $\hat{H}^0\psi_n^0=E_n^0\psi_n^0$.
Here, the subscript labels the wavefunctions and eigenvalues while the
superscript identifies them as solutions to the unperturbed (Newtonian)
Hamiltonian. Because the potential $V(r)=2\pi Gm^2|r|$ is continuous on $%
\mathbb{R}$, non-negative and increases like $|r|$, we know that the system
has isolated eigenvalues $E_n^0$ such that $E_n^0\rightarrow\infty$ as $%
n\rightarrow\infty$ \cite{gusta}. Defining
\begin{equation}
\sigma:=\left(\frac{2\pi G}{\hbar^2}\right)^{\frac{1}{3}}m,
\end{equation}
the normalized wavefunctions are found to
be\footnote{
For the purposes here, an arbitrary phase factor in the normalization
constant is of no importance. See Section 4 for details.  For normalization procedures involving Airy functions see \cite{valle}.}
\begin{equation}
\psi_n^0(r)= \left\{
\begin{array}{lr}
\displaystyle\sqrt{\frac{\sigma}{2a^{\prime}_{\frac{n+2}{2}}}}\frac{1}{%
\left|Ai\left(a^{\prime}_{\frac{n+2}{2}}\right)\right|} Ai\left(\sigma|r|+%
a^{\prime}_{\frac{n+2}{2}}\right), & \hbox{ if } n \in\left\{0\right\}\cup 2%
\mathbb{N}, \smallskip\smallskip \\
\displaystyle\sqrt{\frac{\sigma}{2}}\frac{1}{\left|Ai^{\prime}\left(a_{%
\frac{n+1}{2}}\right)\right|}\hbox{sgn}(r) Ai\left(\sigma|r|+a_{\frac{n+1%
}{2}}\right), & \hbox{ if } n \in 2\mathbb{N}-1 \label{newtWF}%
\end{array}
\right.
\end{equation}
with energies
\begin{equation}
E_n^0= \left\{
\begin{array}{lr}
\displaystyle-\left(2\pi\hbar G\right)^{\frac{2}{3}}ma^{\prime}_{\frac{n+2}{2}},
& \hbox{ if } n \in\left\{0\right\}\cup 2\mathbb{N}, \smallskip\smallskip \\
-\left(2\pi\hbar G\right)^{\frac{2}{3}}ma_{\frac{n+1}{2}}, & \hbox{ if }
n \in 2\mathbb{N}-1%
\end{array}
\right.
\end{equation}
and $Ai(x)$ is the Airy function.

In the above, $a_n$ is the $n^{th}$ root of $Ai(x)$, labeled such that $%
a_{n+1}<a_n$, $n\in\mathbb{N}$, while $a^{\prime}_n$ is the $n^{th}$
root of $Ai^{\prime}(x)$, for which an analogous ordering holds (prime denoting
the derivative with respect to $x$). It should
be noted that for $n\in\mathbb{N}$, $a_n < a^{\prime}_n<0$, so that the
energy eigenvalues are strictly positive and monotonically increasing to
infinity. The states are non-degenerate, and by the orthonormality
relationships $\langle\psi_m^0 \vert \psi_n^0\rangle= \delta_{mn}$ \cite{valle}
and the completeness of $\left\{\psi_n^0\right\}_n$ in $L^2\left(\mathbb{R},%
\mathbb{C}\right)$ \cite{titch}, the eigenfunctions (\ref{newtWF}) form
an orthonormal basis of $L^2\left(\mathbb{R},\mathbb{C}\right)$.

\section{Quantization of the Post-Newtonian Approximation of $(1+1)$%
-Dimensional Dilaton Gravity}

The Hamiltonian (\ref{hamil}) becomes, with the canonical operator
identification in configuration space the Hamiltonian operator $\hat{H}=\hat{H}^{0}+\hat{H}%
^{\prime }$. The spectrum $\left\{ E_{n}^{0}\right\} _{n}$ and
eigenfunctions $\left\{ \psi _{n}^{0}\right\} _{n}$ of the unperturbed
Hamiltonian $\hat{H}^{0}$ were found explicitly above, and it was noted that
the eigenfunctions are non-degenerate and form an orthonormal basis of $%
L^{2}\left( \mathbb{R},\mathbb{C}\right) $. Hence, the Hamiltonian $\hat{H}$
can be solved approximately using non-degenerate perturbation theory \cite%
{ferna}. In the slow moving, weak field limit, we expect the effects of $%
\hat{H}^{\prime }$ to be small as compared to those of $\hat{H}^{0}$. The
validity of this assumption will be discussed in Section 5. \newline
\indent For the total Hamiltonian (\ref{hamil}), the Schr\"{o}dinger equation is $\hat{H}%
\psi _{n}=E_{n}\psi _{n}$. The $n^{th}$ wavefunction $\psi _{n}$ is written
in terms of corrections to the $n^{th}$ Newtonian wavefunction, $\psi
_{n}^{0}$,
\begin{equation*}
\psi _{n}=\psi _{n}^{0}+\sum_{j\geq 1}\psi _{n}^{j}.
\end{equation*}
Here, for $j\in\mathbb{N}$, $\psi_n^j$ is the $j^{th}$ order correction to
the unperturbed wavefunction $\psi_n^0$. The first order correction is
given by \cite{griff}
\begin{equation}
\psi_n^1=\sum_{m\neq n}\frac{\langle\psi_m^0| \hat{H}^{\prime} | \psi_n^0\rangle
}{E_n^0-E_m^0}\psi_m^0  \label{fncorr}
\end{equation}
where $\langle\cdot\vert \cdot\rangle :L^2\left(\mathbb{R},\mathbb{C}\right)
\times L^2\left(\mathbb{R},\mathbb{C}\right) \rightarrow \mathbb{C}$ is the
inner product on the Hilbert space $L^2\left(\mathbb{R},\mathbb{C}\right)$
defined by
\begin{equation*}
\langle f \vert g\rangle:=\int_{\mathbb{R}}\bar{f}g.
\end{equation*}
In the above, $\bar{f}$ denotes the complex conjugate of $f$.

The energy eigenvalues are written in an analogous form,
\begin{equation*}
E_n=E_n^0+\sum_{j\geq1}E_n^j
\end{equation*}
with $E_n^j$ defined as the $j^{th}$ order correction to the unperturbed
energy eigenvalue $E_n^0$. The first order correction to the energy
eigenvalues is given by \cite{griff}
\begin{equation}
E_n^1=\langle\psi_n^0 \vert \hat{H}^{\prime}\vert \psi_n^0\rangle.
\end{equation}
We are now in a position to calculate the energy and eigenfunction
corrections. \newline
\indent Explicitly, the Hamiltonian operator for the perturbation, $\hat{H}%
^{\prime}$, acting on an unperturbed state $\psi_n^0$ is
\begin{equation}
\hat{H}^{\prime}\psi_n^0=-\frac{\hbar^4}{4m^3c^2}\partial_r^4\psi_n^0-\frac{%
4\pi\hbar^2 G}{c^2}|r|\partial_r^2\psi_n^0
\end{equation}
where we have used the so-called natural ordering of operators. Observe that
if $m$ is odd and $n$ is even (or vice-versa), we have $\langle\psi_m^0 \vert \hat{H}%
^{\prime}\vert \psi_n^0\rangle=0$. Indeed, if $n$ is zero or even, then $\psi_n^0$
is even, and hence so too is $\partial_r^{2k}\psi_n^0$, $k\in\mathbb{N}$.
Then, the inner products $\langle\psi_m^0 \vert \partial_r^{2k}\vert \psi_n^0\rangle$
vanish, as they are integrals of an odd function over an even domain. From
the definition of $\psi_n^1$, we see that this implies that the parity of a
state is preserved by the perturbations, as expected since the
Hamiltonian (\ref{hamil}) has parity symmetry. Hence only when $m$ and $n$ are both odd (or both even) may
$\langle\psi_m^0 \vert \hat{H}^{\prime}\vert \psi_n^0\rangle$ be non-vanishing.

We now introduce a useful property of the Airy function that allows
the explicit calculation of the desired inner products. It follows from the
identity $Ai^{\prime\prime}(x)=xAi(x)$ that we can write $%
Ai^{(n)}(x)=f_n(x)Ai(x)+g_n(x)Ai^{\prime}(x)$, where $f_n(x)$ and $g_n(x)$
satisfy the recursive relationships
\begin{equation}
f_{n+1}(x)=f_n^{\prime}(x)+xg_n(x),\;\;\;
g_{n+1}(x)=f_n(x)+g_n^{\prime}(x),\;\;\; \forall n\in\mathbb{N}
\end{equation}
with $f_0(x)=1$ and $g_0(x)=0$. For example, using the definitions of $f_n$
and $g_n$, we get
\begin{equation}
\partial_r^4\psi_n^0=\sigma^6r^2\psi_n^0+2\sigma^5\delta_n|r|\psi_n^0+%
\sigma^4\delta_n^2\psi_n^0+2\sigma^4\psi_n^{0\prime}
\end{equation}
where $\delta_n$ is one of $a_n$ or $a^{\prime}_n$.

We therefore write
for $m,n\in\mathbb{N}$,
\begin{equation}
\begin{array}{lcl}
\displaystyle \langle\psi_m^0 \vert \hat{H}^{\prime}\vert \psi_n^0\rangle & = & %
\displaystyle -\left(\frac{\hbar^4}{4m^3c^2}\sigma^6+\frac{4\pi\hbar^2 G}{c^2%
}\sigma^3\right)\langle\psi_m^0 \vert r^2\vert \psi_n^0\rangle \\
&  &  \\
&  & -\displaystyle\left(\frac{\hbar^4}{2m^3c^2}\sigma^5\delta_n+\frac{%
4\pi\hbar^2 G}{c^2}\sigma^2\delta_n\right)\langle\psi_m^0 \vert  |r| \vert \psi_n^0\rangle
\\
&  &  \\
&  & -\displaystyle\frac{\hbar^4}{4m^3c^2}\sigma^4\delta_n^2\langle\psi_m^0 \vert %
\psi_n^0\rangle-\frac{\hbar^4}{2m^3c^2}\sigma^4\langle\psi_m^0 \vert\psi_n^{0\prime}\rangle. \label{innprod}%
\end{array}%
\end{equation}
We evaluate the above expression with $m=n$, which gives $E_n^1$. All inner
products can be evaluated explicitly.\footnote{%
See Appendix.} After putting in the explicit form of $\sigma$, we obtain
\begin{equation}
E_n^1= \left\{
\begin{array}{lr}
\frac{m}{2}\frac{\left(2\pi\hbar G\right)^{\frac{4}{3}}}{c^2}\left(\frac{13}{%
30}\left(a_{\frac{n+2}{2}}^{{\prime}}\right)^2+\frac{9}{10}\frac{1}{a^{\prime}_{\frac{n+2}{2}}}%
\right) & \hbox{ if } n \in\left\{0\right\}\cup 2\mathbb{N}, \\
\frac{m}{2}\frac{\left(2\pi\hbar G\right)^{\frac{4}{3}}}{c^2}\left(\frac{13}{%
30}a_{\frac{n+1}{2}}^2\right) & \hbox{ if } n \in 2\mathbb{N}-1.%
\end{array}
\right.
\end{equation}

Inserting appropriate values of $a_n$ and $a^{\prime}_n$ shows that $E_0^0<0$
while $E_n^0>0$ for $n\geq1$. By the ordering and unboundedness of $a_n$
and $a^{\prime}_n$, $E_n^1$ is monotonically increasing in $n$ without bound,
and so the approximation will be valid for only finitely many
states, a point we shall return to in Section 5. The magnitude of the energy corrections is expected to be
quite small,
as is common in quantum gravity. Indeed, for the ground state, we have
\begin{equation}\label{enratio}
\left| \frac{E_0^1}{E_0^0}\right| \approx 0.21\frac{\left(2\pi\hbar
G\right)^{\frac{2}{3}}}{c^2},
\end{equation}
where the lower dimensionality implies that $G$ has the same dimensions as $c^3/\hbar$.

Proceeding, we can now find expressions for the perturbed wavefunctions. To
do so, we evaluate (\ref{innprod}) with $m$ and $n$ both zero or even to get
\begin{equation}
\langle\psi_m^0 \vert \hat{H}^{\prime}\vert \psi_n^0\rangle=\frac{m}{2}\frac{%
\left(2\pi\hbar G\right)^{\frac{4}{3}}}{c^2}\left[54+5a^{\prime}_{\frac{n+2}{2}%
}\left(a^{\prime}_{\frac{m+2}{2}}-a^{\prime}_{\frac{n+2}{2}}\right)^2\right]\frac{%
(-1)^{m+n+1}\left(a^{\prime}_{\frac{m+2}{2}}+a^{\prime}_{\frac{n+2}{2}}\right)}{%
\left(a^{\prime}_{\frac{m+2}{2}}-a^{\prime}_{\frac{n+2}{2}}\right)^4\sqrt{a^{\prime}_{%
\frac{m+2}{2}}a^{\prime}_{\frac{n+2}{2}}}}  \label{evenProd}
\end{equation}
and for $m$ and $n$ both odd,
\begin{equation}
\langle\psi_m^0 \vert \hat{H}^{\prime}\vert \psi_n^0\rangle=\frac{m}{2}\frac{%
\left(2\pi\hbar G\right)^{\frac{4}{3}}}{c^2}\left[54+5a_{\frac{n+1}{2}%
}\left(a_{\frac{m+1}{2}}-a_{\frac{n+1}{2}}\right)^2\right]\frac{%
(-1)^{m+n}2}{\left(a_{\frac{m+1}{2}}-a_{\frac{n+1}{2}}\right)^4}.
\label{oddProd}
\end{equation}
Using Eqs. (\ref{fncorr}), (\ref{evenProd}) and (\ref{oddProd}), we have
\begin{equation}
\psi_n^1=\sum_{m\geq0}\alpha_{mn}\psi_m^0
\end{equation}
where
\begin{equation}
\alpha_{mn}= \left\{
\begin{array}{lr}
\frac{\left(2\pi\hbar G\right)^{\frac{2}{3}}}{2c^2}\left(-1\right)^{m+n+1}
\frac{\left(54+5a^{\prime}_{\frac{n+2}{2}}\left(a^{\prime}_{\frac{m+2}{2}}- a^{\prime}_{%
\frac{n+2}{2}}\right)^2\right)\left(a^{\prime}_{\frac{m+2}{2}}+a^{\prime}_{\frac{n+2%
}{2}}\right)}{\left(a^{\prime}_{\frac{m+2}{2}}-a^{\prime}_{\frac{n+2}{2}}\right)^5%
\sqrt{a^{\prime}_{\frac{m+2}{2}}a^{\prime}_{\frac{n+2}{2}}}}, & \hbox{ if } m\neq n,
m,n \in\left\{0\right\}\cup 2\mathbb{N} \bigskip \\
\frac{\left(2\pi\hbar G\right)^{\frac{2}{3}}}{c^2}(-1)^{m+n}\frac{%
\left(54+5a_{\frac{n+1}{2}}\left(a_{\frac{m+1}{2}}-a_{\frac{n+1}{%
2}}\right)^2\right)}{\left(a_{\frac{m+1}{2}}-a_{\frac{n+1}{2}%
}\right)^5}, & \hbox{ if } m\neq n, m,n \in 2\mathbb{N}-1 \\
0, & \mbox{ otherwise.} \label{alphaMN}%
\end{array}
\right.
\end{equation}

\begin{figure}[t]
\psfrag{psi}{$\psi (r)$}
\psfrag{r}{$r$}
\includegraphics[scale=0.25]{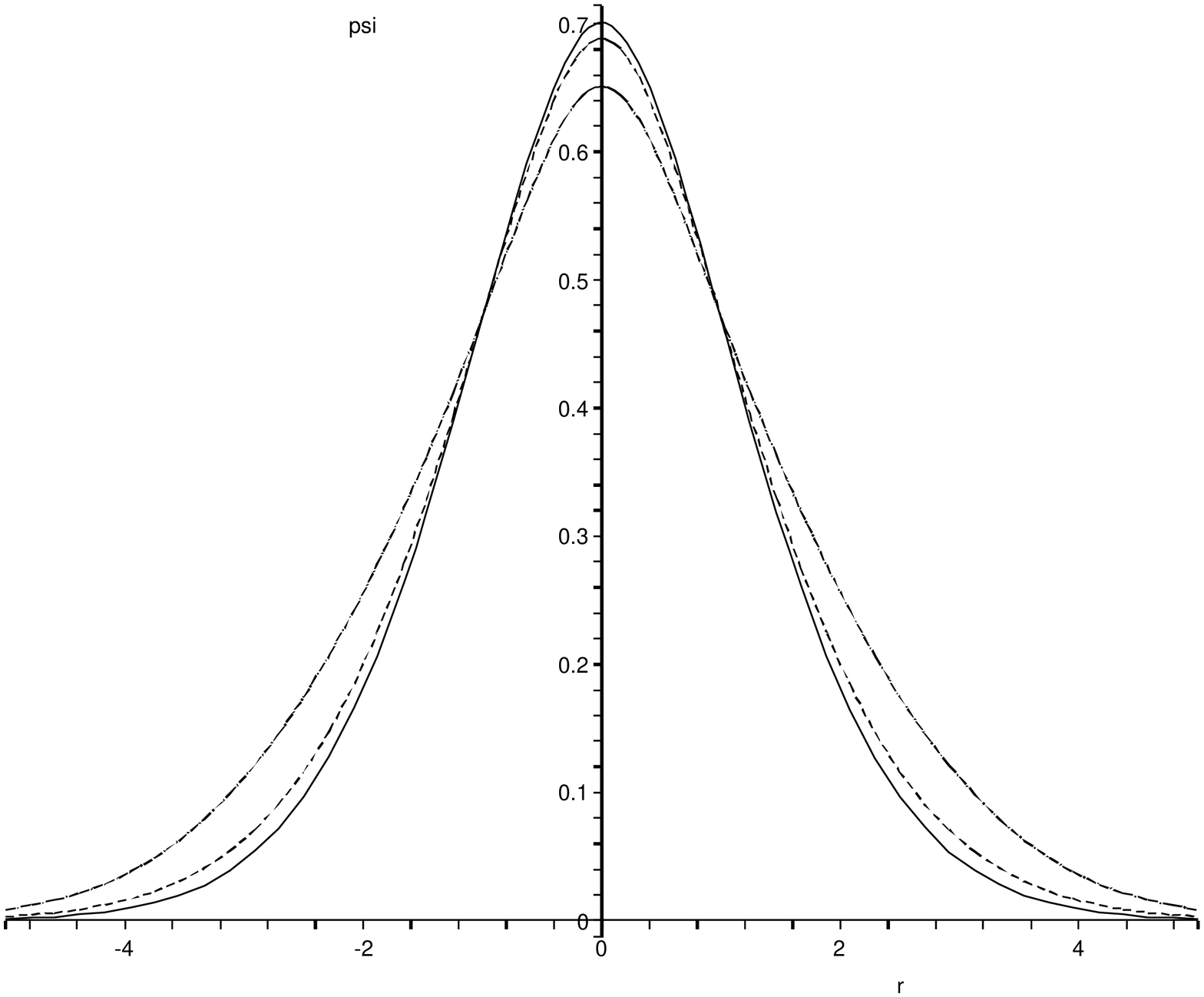}
\psfrag{psi^2}{$\vert \psi (r) \vert ^2$}
\psfrag{r}{$r$}
\includegraphics[scale=0.25]{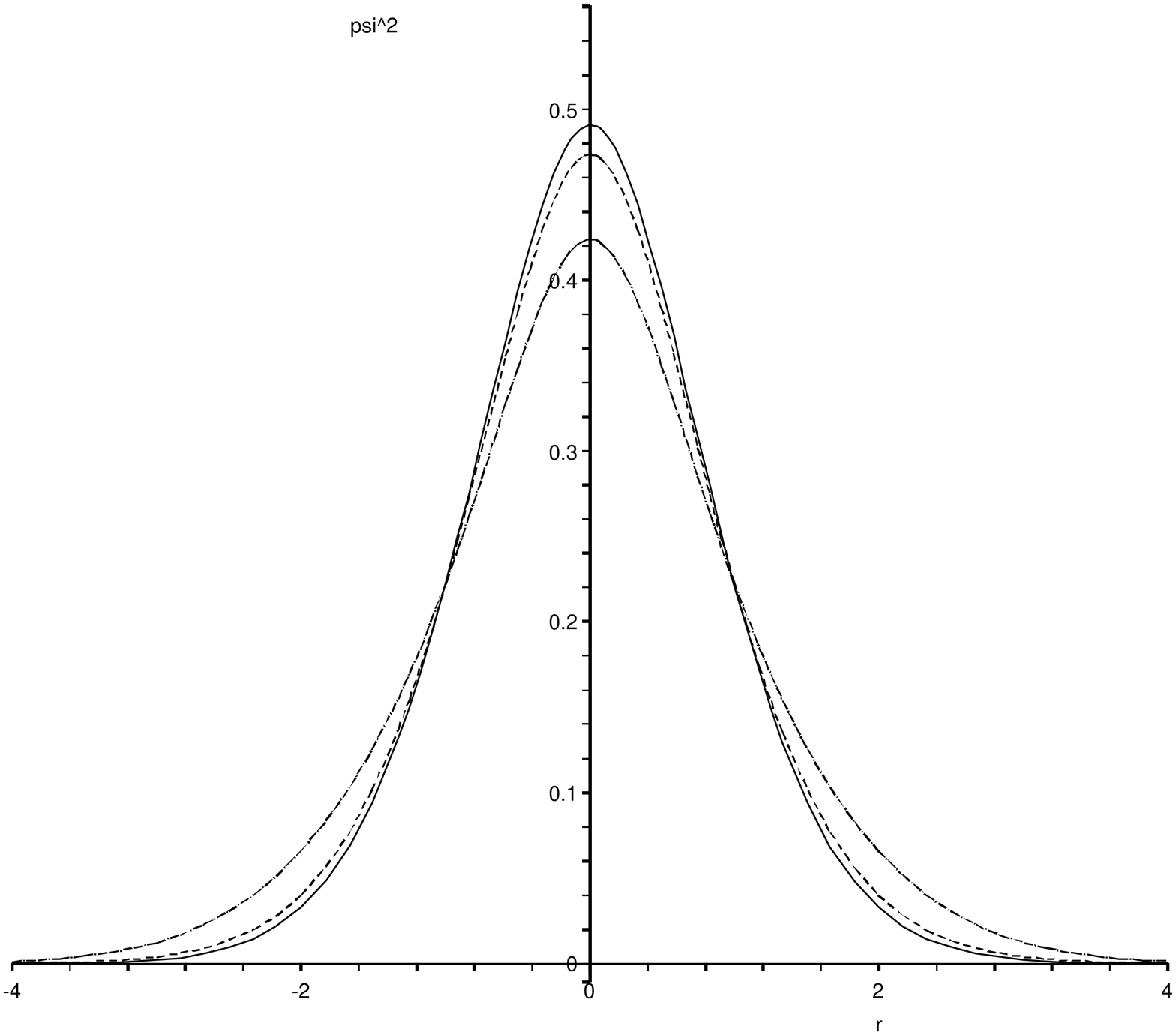}
\caption{The figure to the left shows a plot of $\protect\psi_0^0$ $\left(
\hbox{solid}\right)$, $\protect\psi_0$ for $c=7$ $\left(\hbox{dash}\right)$
and $\protect\psi_0$ for $c=3$ $\left(\hbox{dash-dot}\right)$. On the right,
the plot shows $\left|\protect\psi_0^0\right|^2$ $\left(\hbox{solid}\right)$%
, $\left|\protect\psi_0\right|^2$ for $c=7$ $\left(\hbox{dash}\right)$ and $%
\left|\protect\psi_0\right|^2$ for $c=3$ $\left(\hbox{dash-dot}\right)$.
Units are chosen such that $\hbar=1 J\cdot s$, $G=\frac{1}{2\protect\pi} \frac{m}{kg\cdot s^2}$, and $m=1 kg$. }
\label{wavefunctions}
\end{figure}

The first order perturbed wavefunctions are thus given by\footnote{%
Had a phase factor been included in $\psi_n^0$, say $\tilde{\psi}%
_n^0=e^{i\theta_n}\psi_n^0$ for some $\theta_n\in\mathbb{R}$, then we would
now have that $\tilde{\psi}_n=e^{i\theta_n}\psi_n$. For simplicity
we omit this phase factor. }
\begin{equation}
\psi_n=\psi_n^0+\sum_{m\geq0}\alpha_{mn}\psi_m^0.\;\;\;
\end{equation}
It is easy to check that to second order in $c^{-1}$, $\psi_n$ is
normalized. See Figure \ref{wavefunctions} for a plot of the perturbed
wavefunctions and probability densities for increasing values of $c$.

\section{Expectation Values and the Validity of the Corrections}

We now wish to calculate the expectation values of powers of $\hat{r}$ and $\hat{p}$ so
as to compare the Newtonian and post-Newtonian systems, and to check the limits of validity of our perturbation results. First, some
notation. We define, for $\phi\in D(\hat{Q})\subset L^2\left(\mathbb{R},%
\mathbb{C}\right)$,
\begin{equation}
\langle\hat{Q}\rangle_{\phi}:=\langle{\phi \vert \hat{Q} \vert \phi\rangle}
\end{equation}
for some observable $\hat{Q}$.

We state results for the perturbed wavefunctions, which are derived using the inner product identities developed in the Appendix.  Taking the $c\rightarrow\infty$ limit, these also give the Newtonian expectation values.
We have, keeping only terms to second order in $c^{-1}$,
\begin{equation}
\langle\hat{r}\rangle_{\psi_n}=0, \;\;\;\;  \langle\hat{p}\rangle_{\psi_n}=0, \qquad
\end{equation}
reflecting the fact that the perturbation (\ref{pertHam}) is parity preserving.  We also find
\begin{equation}
\langle\hat{r}^2\rangle_{\psi_n}=\left\{
\begin{array}{ll}
\begin{array}{l}
\displaystyle\left[\frac{8a_{\frac{n+2}{2}}^{\prime 3}-3}{15a^{\prime}_{%
\frac{n+2}{2}}}+24\displaystyle\sum_{m\geq0}\alpha_{mn}\frac{(-1)^{m+n}\left(%
a^{\prime}_{\frac{m+2}{2}}+ a^{\prime}_{\frac{n+2}{2}}\right)}{\left(a^{\prime}_{\frac{%
n+2}{2}}-a^{\prime}_{\frac{m+2}{2}}\right)^4\sqrt{a^{\prime}_{\frac{m+2}{2}}a^{\prime}_{%
\frac{n+2}{2}}}}\right]\left(\frac{\hbar^2}{2\pi G}\right)^{\frac{2}{3}}\frac{1}{m^2}%
\end{array}
, & \hbox{ if } n \in\left\{0\right\}\cup 2\mathbb{N} \bigskip\bigskip \\
\begin{array}{l}
\displaystyle\displaystyle\left[\frac{8a_{\frac{n+1}{2}}^2}{15}%
+48\sum_{m\geq0}\alpha_{mn}\frac{(-1)^{m+n+1}}{\left(a_{\frac{n+1}{2}}-a_{%
\frac{m+1}{2}}\right)^4}\right]\left(\frac{\hbar^2}{2\pi G}\right)^{\frac{2}{%
3}}\frac{1}{m^2}%
\end{array}%
, & \hbox{ if } n \in 2\mathbb{N}-1,%
\end{array}
\right.
\end{equation}
and
\begin{equation}
\langle\hat{p}^2\rangle_{\psi_n}=\left\{
\begin{array}{ll}
\begin{array}{l}
-\left[\frac{1}{3}a^{\prime}_{\frac{n+2}{2}}+2\displaystyle\sum_{m\geq0}\alpha_{mn}
\frac{(-1)^{m+n}\left(a^{\prime}_{\frac{m+2}{2}}+ a^{\prime}_{\frac{n+2}{2}}\right)}{%
\left(a^{\prime}_{\frac{m+2}{2}}-a^{\prime}_{\frac{n+2}{2}}\right)^2\sqrt{a^{\prime}_{%
\frac{m+2}{2}}a^{\prime}_{\frac{n+2}{2}}}}\right]\left(2\pi\hbar G\right)^{\frac{%
2}{3}}m^2%
\end{array}%
, & \hbox{ if } n \in\left\{0\right\}\cup 2\mathbb{N} \bigskip\bigskip \\
\begin{array}{l}
-\left[\frac{1}{3}a_{\frac{n+1}{2}}+ 2\displaystyle\sum_{m\geq0}\alpha_{mn}
\frac{(-1)^{m+n+1}}{\left(a_{\frac{m+1}{2}}-a_{\frac{n+1}{2}%
}\right)^2}\right]\left(2\pi\hbar G\right)^{\frac{2}{3}}m^2%
\end{array}%
, & \hbox{ if } n \in 2\mathbb{N}-1.%
\end{array}
\right.
\end{equation}

We see that $\langle\hat{r}^2\rangle_{\psi_n}$ and $\langle\hat{p}^2\rangle_{\psi_n}$ depend strongly on the masses of the particles, being proportional to the inverse square and square of the mass, respectively.  This dependence on mass resembles an exaggerated version of the simple harmonic oscillator (SHO), with $\langle\hat{r}^2\rangle^{SHO}$ and $\langle\hat{p}^2\rangle^{SHO}$ being proportional to the inverse root and root of the mass, respectively.  For particles of small (large) mass, the square separation quickly becomes very large (small) and changes slowly (quickly) due to the $m^2$ dependence of $\langle\hat{p}^2\rangle_{\psi_n}$.It is also instructive to compare these results to those of $(3+1)$-dimensional Newtonian gravity.  In the $(3+1)$ case, $\langle\hat{r}^2\rangle$ and $\langle\hat{p}^2\rangle$ are proportional to $m^{-6}$ and $m^6$, respectively.  In this sense  the expectation values are an even more exaggerated version of the SHO values.

\begin{figure}[t]
\psfrag{r}{$\langle \hat{r}^2 \rangle$}
\psfrag{c}{$c$}
\psfrag{n=0}{$n=0$}
\psfrag{n=1}{$n=1$}
\includegraphics[scale=0.25]{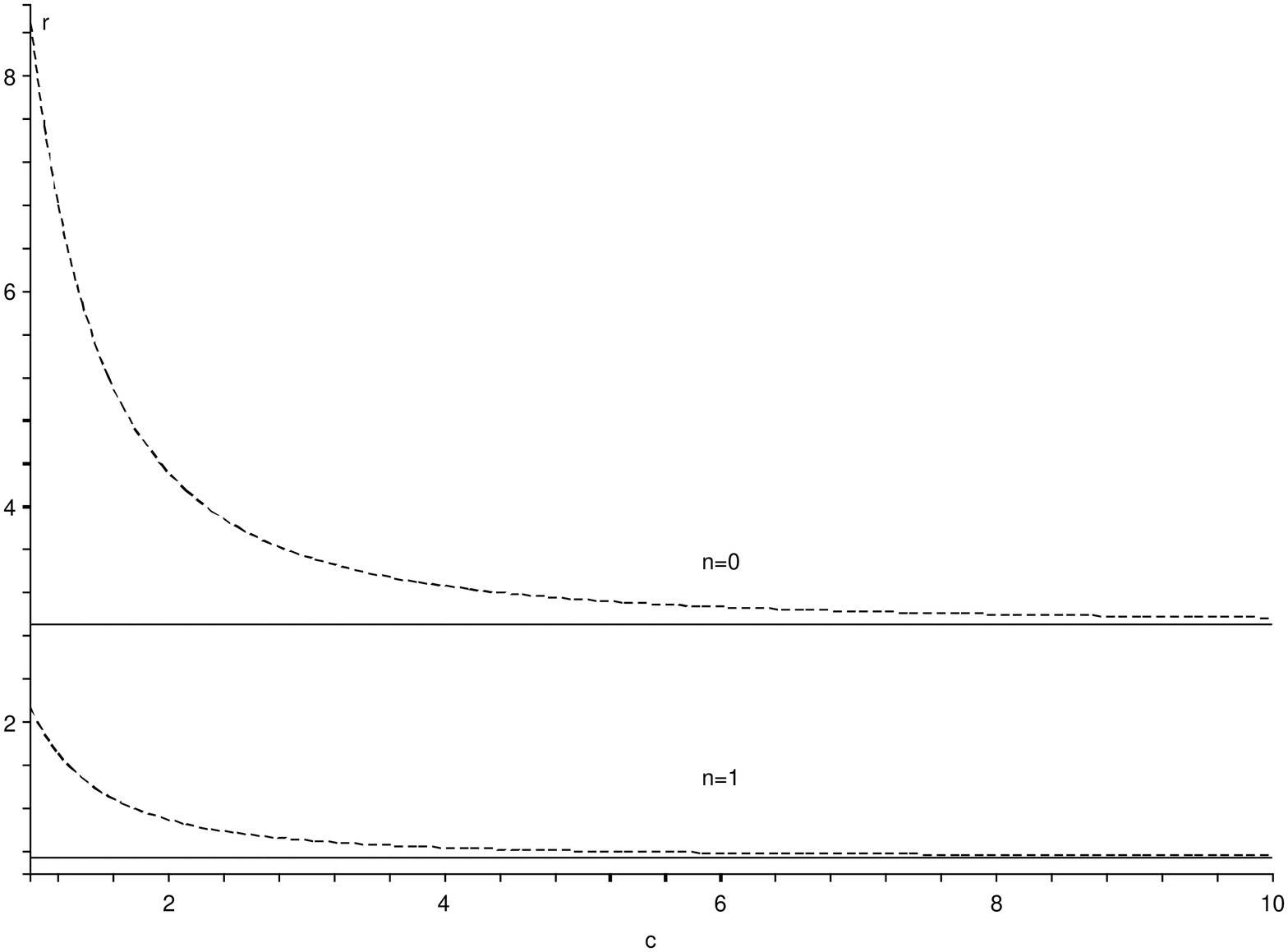}
\psfrag{p}{$\langle \hat{p}^2 \rangle$}
\psfrag{c}{$c$}
\psfrag{n=0}{$n=0$}
\psfrag{n=1}{$n=1$}
\includegraphics[scale=0.25]{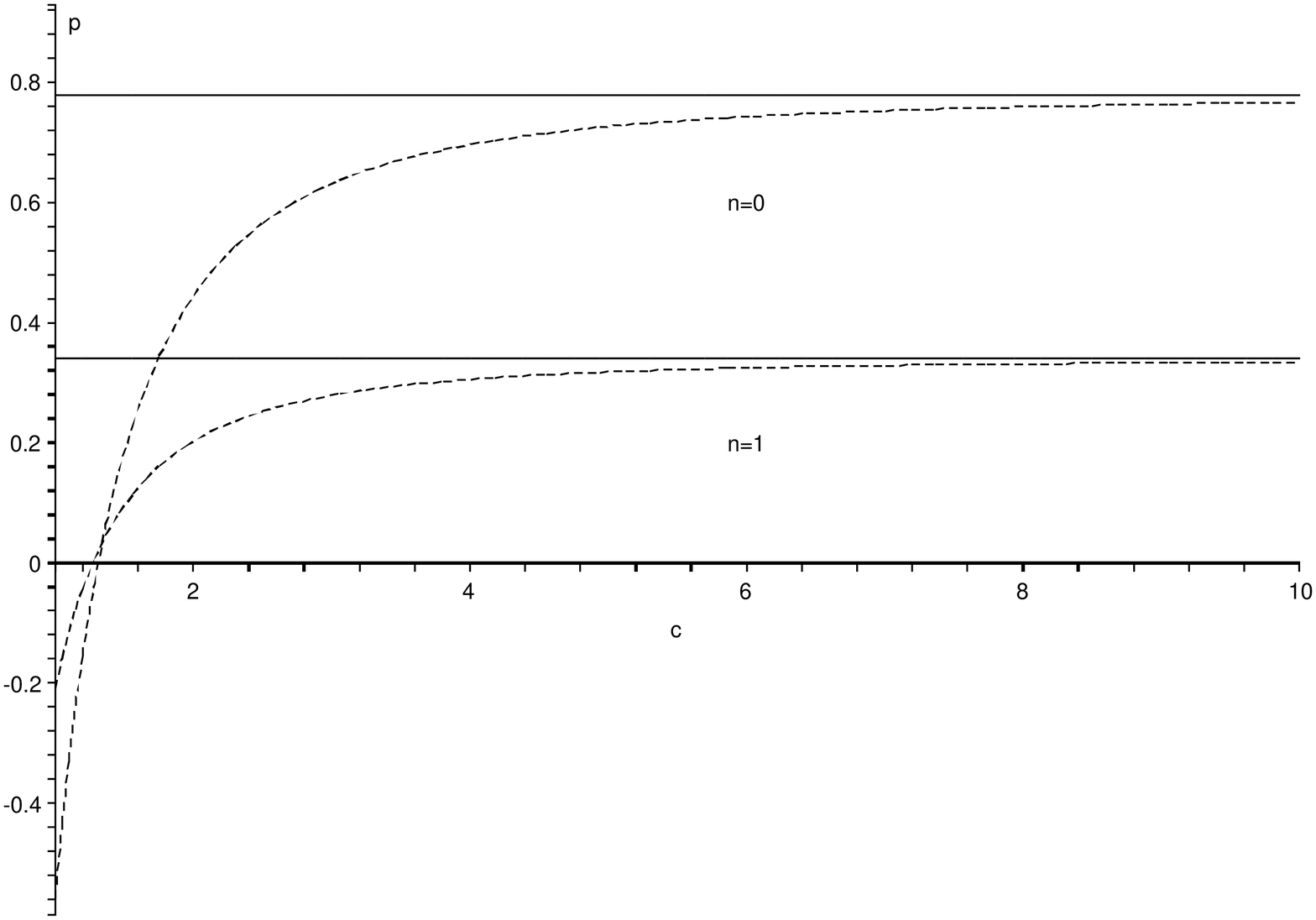}
\caption{The figure to the left shows a plots, for $n=0,1$, of $\langle\hat{r}^2\rangle_{\psi_n}$ (dot) as a function of $c$ and $\langle\hat{r}^2\rangle_{\psi_n^0}$ (solid), illustrating $\lim_{c\rightarrow \infty}\langle\hat{r}^2\rangle_{\psi_n} = \langle\hat{r}^2\rangle_{\psi_n^0}$. To the right, plots of $\langle\hat{p}%
^2\rangle_{\protect\psi_n}$, $n=0,1$, (dot) as a function of $c$ and $\langle\hat{p}%
^2\rangle_{\protect\psi_n^0}$ (solid) are shown, illustrating $\lim_{c\rightarrow \infty}\langle\hat{p}^2\rangle_{\psi_n} = \langle\hat{p}^2\rangle_{\psi_n^0}$.  Units are chosen such that $\hbar=1 J\cdot s$, $G=\frac{1}{2\protect\pi} \frac{m}{kg\cdot s^2}$, and $m=1 kg$. }
\label{expVals}
\end{figure}

From Eq. (\ref{alphaMN}), we see that $\displaystyle\lim_{c\rightarrow\infty}\alpha_{mn}=0$.  Using this, it is straightforward to check that we have $\displaystyle\lim_{c\rightarrow \infty}\langle\hat{Q}\rangle_{\psi_n}=
\langle\hat{Q}\rangle_{\psi_n^0}$ for $\hat{Q}=\hat{r}^n, \hat{p}^n$, $n=1,2$, as claimed above.  This is illustrated in Figure \ref{expVals}.  From the expression for $\alpha_{mn}$, we see that the expectation values diverge as $c\rightarrow0$.  This, however, is not of concern as the approximation is valid only for large $c$. Higher order contributions
in $c^{-1}$ will greatly modify Figure \ref{expVals} for small $c$, while
leaving the behaviour mostly unchanged for large $c$.

As an example, in units such that $\hbar=1$, $G=\frac{1}{2\protect\pi} $, and $m=1$, we have $\langle\hat{r}^2\rangle_{\psi_0^0}%
\approx0.7499$ and $\langle\hat{p}^2\rangle_{\psi_0^0}\approx0.3396$. Setting
$c=10$, and including twenty terms (i.e. $m=0,\dots,19$) in the sums, we
get $\langle\hat{r}^2\rangle_{\psi_0}\approx0.7636$ and $\langle\hat{p}%
^2\rangle_{\psi_0}\approx0.3341$.

\begin{figure}[t]
\psfrag{unc}{$(\Delta \hat{x})(\Delta \hat{p})$}
\psfrag{c}{$c$}
\psfrag{hbar/2}{$\frac{\hbar}{2}$}
\psfrag{newt}{$(\Delta \hat{x})_{\psi^0_0}(\Delta \hat{p})_{\psi^0_0}$}
\psfrag{quant}{$(\Delta \hat{x})_{\psi_0}(\Delta \hat{p})_{\psi_0}$}
\includegraphics[scale=0.30]{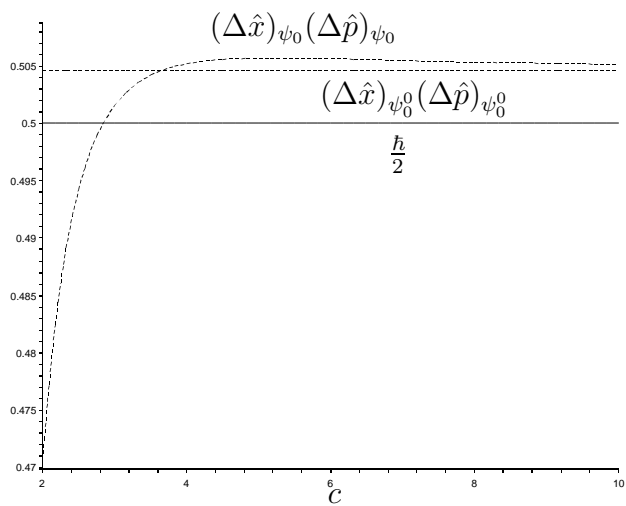}
\caption{The figure shows a plot  $\frac{\hbar}{2}$ (solid), $(\Delta \hat{x})_{\psi_0^0}(\Delta \hat{p})_{\psi_0^0}$ and $(\Delta \hat{x})_{\psi_0}(\Delta \hat{p})_{\psi_0}$ (both dash) as a function of $c$, illustrating $\lim_{c\rightarrow\infty} (\Delta \hat{x})_{\psi_0}(\Delta \hat{p})_{\psi_0}=(\Delta \hat{x})_{\psi_0^0}(\Delta \hat{p})_{\psi_0^0}$. Units are such that $\hbar=1 J\cdot s$, $G=\frac{1}{2\protect\pi} \frac{m}{kg\cdot s^2}$, and $m=1 kg$. }
\label{uncertPlot}
\end{figure}

Denoting the dispersion of an
observable $\hat{Q}$ in the state $\phi$ by $(\Delta \hat{Q})_{\phi}$, using the above expectation
values, we see that
\begin{equation}
\left(\Delta \hat{x}\right)_{\psi_n^0} \left(\Delta \hat{p}\right)_{\psi_n^0} =\left\{
\begin{array}{ll}
\frac{1}{3}\sqrt{\frac{3-8\left(a_{\frac{n+2}{2}}^{\prime}\right)^3}{5}}\hbar, & \hbox{ if }
n \in\left\{0\right\}\cup 2\mathbb{N} \\
\frac{2}{3}\sqrt{\frac{2}{5}}\vert a_{\frac{n+1}{2}}\vert\hbar, & \hbox{
if } n \in 2\mathbb{N}-1.%
\end{array}
\right.
\end{equation}
It is straightforward to show that the uncertainty
principle is not violated upon substituting values for $a_n$ and $a^{\prime}_n$.
Figure \ref{uncertPlot} shows a plot of $\left(\Delta \hat{x}\right) \left(\Delta \hat{p}\right)$ for both the perturbed and unperturbed ground state wavefunctions, as well as the lower bound $\frac{\hbar}{2}$ set by the uncertainty principle.
Of course for small enough $c$ we find that $\left(\Delta \hat{x}\right)_{\psi_0^0} \left(\Delta \hat{p}\right)_{\psi_0^0}<\frac{\hbar}{2}$, indicative of the limits of the validity of our approximation.

 To obtain an estimate of the number of states for which our approximation holds, we demand that the expected gravitational binding energy be less than the rest energy of the system \cite{weinb}:
\[
2\pi Gm^2|r|<2mc^2.
\]
Quantum mechanically,  this suggests the bound
\[
\langle \hat{r}^2\rangle_{\psi_n}<\frac{c^4}{\pi^2 G^2m^2}
\]
which is a bound on $n$, since it  requires
\begin{equation}
\frac{8a_{\frac{n+1}{2}}^2}{15}%
+48\sum_{m\geq0}\alpha_{mn}\frac{(-1)^{m+n+1}}{\left(a_{\frac{n+1}{2}}-a_{%
\frac{m+1}{2}}\right)^4} < \frac{4c^4}{\left(2\pi\hbar G\right)^{\frac{4}{3}}}.
\end{equation}
For $\hbar=1$, $G=\frac{1}{2\pi}$, $c=10$, this gives $n\lesssim645$.  So, although the number of states is finite, even for small $c$, this number is quite large.

\section{Discussion}

As noted
above,  the Hamiltonian (\ref{exactHam}) represents the exact energy functional of all
of the degrees of freedom in the relativistic 2-body system. Hence its quantization
is tantamount to a full quantization of gravity coupled to matter in (1+1)-dimensions.

We have successfully completed the first steps to this end.
We have used perturbation theory to obtain explicit energy and eigenfunction corrections to the relativistically perturbed
Newtonian Hamiltonian to second order in $c^{-1}$.
We found that the ground state energy in the
relativistic case is lower than in the Newtonian case, while all other
relativistic states have higher energies than their Newtonian counterparts.
As suggested in Figure \ref{expVals}, we  conjecture that $\langle\hat{r}%
^2\rangle_{\psi_n^0} < \langle\hat{r}^2\rangle_{\psi_n}$ and $\langle\hat{p}%
^2\rangle_{\psi_n^0} > \langle\hat{p}^2\rangle_{\psi_n}$. That is,
relativistically, the two particles are not bound as tightly together as in
Newtonian theory.  The above inequalities have been verified numerically for $n\leq 20$,
including forty terms in the sums.  On the other hand, on average we expect the separation of
the particles to be changing more slowly in the relativistic case than in its
Newtonian counterpart.

In estimating the actual magnitude of the quantum-gravitational corrections to the Newtonian
system, we run into the problem of not knowing
the appropriate values of the fundamental constants
$c$, $\hbar$ and $G$ in the $(1+1)$-dimensional case. However if we view the $R=T$ theory as
being the effective relativistic theory near the surface of the earth we can
obtain a numerical value of the energy correction.  The $(3+1)$ Newtonian potential close to the surface of the Earth is
\[
V^{(4)}(r)=-\frac{G^{(4)}M_{\oplus}m}{\vert R_{\oplus}+r\vert}.
\]
For small $r$, this becomes
\[
V^{(4)}(r)=-\frac{G^{(4)}M_{\oplus}m}{R_{\oplus}}+\frac{G^{(4)}M_{\oplus}m}{R_{\oplus}^2}r+\mathcal{O}(r^2).
\]
On the other hand, the $(1+1)$ Newtonian potential for the two masses is
\[
V^{(2)}(r)=2\pi GM_{\oplus}m|r|.
\]
Here the superscripts in parentheses refer to the spacetime dimension in which the quantity is valid
(where $G:=G^{(2)}$).  Ignoring the constant term, comparison of the two cases suggests
\begin{equation}
G=\frac{G^{(4)}}{2\pi R_{\oplus}^2}
\end{equation}
which indeed provides $G$ with the correct units.  Using this, with the usual values of $\hbar$ and $c$, we  find $\left| \frac{E_0^1}{E_0^0}\right| \approx 8\times 10^{-56}$.

Of course the actual theory correcting Newtonian gravity in $(3+1)$ dimensions is general relativity and not
the $R=T$ theory.  However the results obtained here illustrate -- quantitatively and qualitatively --  that
sensible quantum-gravitational corrections to Newtonian gravity can be meaningfully computed,  though
the suppression of the effective 2-dimensional Newton constant $G$ renders such effects unobservable
in practice.  A full quantization
of the $R=T$ theory remains an interesting problem for future study.

\section{Appendix}

\subsection{Inner products of the form $\langle\protect\psi_m^0 \vert|r|^i \vert \protect%
\psi_n^0\rangle$}

In calculating inner products involving $\psi_n^0$, we are led to consider
integrals of the form $\int_0^{\infty}dr\,|r|^iAi(r+\delta_1)Ai(r+\delta_2)$%
, $i=1,2$ and $\delta_j\in\mathbb{R}$. Such integrals are found in \cite%
{valle}. We must investigate the cases in which $\delta_1=\delta_2$ and $%
\delta_1\neq\delta_2$ separately.

\subsubsection{$\protect\delta_1=\protect\delta_2$}

We will show explicit calculations for the case $q=1$, which illustrates the
general procedure, and state the remaining cases as results. From \cite%
{valle},
\begin{equation*}
\begin{array}{lcl}
\displaystyle\int dx\,xAi^2\left(x+\delta\right) & = & \displaystyle\frac{1}{%
3}\left(x^2-x\delta-2\delta^2\right)Ai^2\left(x+\delta\right)+\displaystyle%
\frac{2\delta-x}{3}Ai^{\prime 2}\left(x+\delta\right) \\
&  &  \\
&  & +\displaystyle\frac{1}{3}Ai^{\prime}\left(x+\delta\right))Ai\left(x+%
\delta\right).%
\end{array}
\end{equation*}
Using this result for $\psi_n^0$, with, say, $n$ odd, we get after setting $%
\rho:=\sigma r$,
\begin{equation*}
\langle\psi_n^0 \vert|r| \vert\psi_n^0\rangle=\frac{2a_{\frac{n+1}{2}}^2}{\sigma^2}%
\int_0^{\infty}d\rho\,\rho Ai^2\left(\rho+a_{\frac{n+1}{2}}\right)=-%
\frac{4a_{\frac{n+1}{2}}^2}{3\sigma^2}Ai^{\prime 2}\left(a_{\frac{%
n+1}{2}}\right).
\end{equation*}
Using the expression for $a_n$, we see get the final result\footnote{%
We used the fact that $Ai(x)\sim \frac{1}{2\sqrt{\pi}}\frac{e^{-\frac{2}{3}%
x^{\frac{3}{2}}}}{x^{\frac{1}{4}}}$ as $x\rightarrow \infty$ \cite{abram},
so that $x^qAi(x)\rightarrow 0$ as $x\rightarrow \infty$ $\forall q\in%
\mathbb{N}$.}
\begin{equation*}
\langle\psi_n^0 \vert |r| \vert \psi_n^0\rangle=-\frac{2a_{\frac{n+1}{2}}}{3\sigma}.
\end{equation*}
Similarly, for $n$ even, we get
\begin{equation*}
\langle\psi_n^0\vert |r| \vert\psi_n^0\rangle=-\frac{2a^{\prime}_{\frac{n+2}{2}}}{3\sigma}.
\end{equation*}
For $n$ odd,
\begin{equation*}
\langle\psi_n^0\vert r^2 \vert \psi_n^0\rangle=\frac{8a_{\frac{n+1}{2}}^2}{15\sigma^2%
},
\end{equation*}
while for $n$ even,
\begin{equation*}
\langle\psi_n^0 \vert r^2 \vert\psi_n^0\rangle=\frac{8a_{\frac{n+2}{2}}^{\prime 3}-3}{%
15\sigma^2a^{\prime}_{\frac{n+2}{2}}}.
\end{equation*}

\subsubsection{$\protect\delta_1\neq\protect\delta_2$}

Again, an explicit calculation will be shown for the case $q=1$ case, while
the remaining cases will be stated without derivations. From \cite{valle},
\begin{equation*}
\begin{array}{rcl}
\displaystyle\int dx\,xAi\left(x+\delta_1\right)Ai\left(x+\delta_2\right) & =
& \displaystyle -\frac{\delta_1+\delta_2+2x}{\left(\delta_1-\delta_2\right)^2%
}Ai\left(x+\delta_1\right)Ai\left(x+\delta_2\right) \\
&  &  \\
&  & +\displaystyle\left[\frac{x}{\delta_1-\delta_2}+\frac{2}{%
\left(\delta_1-\delta_2\right)^3}\right]\left\{Ai\left(x+\delta_1\right)Ai^{%
\prime}\left(x+\delta_2\right)\right. \\
&  &  \\
&  & \displaystyle\left.-Ai^{\prime}\left(x+\delta_1\right)Ai\left(x+%
\delta_2\right)\right\} \\
&  &  \\
&  & +\displaystyle\frac{2}{\left(\delta_1-\delta_2\right)^2}%
Ai^{\prime}\left(x+\delta_1\right)Ai^{\prime}\left(x+\delta_2\right).%
\end{array}
\end{equation*}
Using this result for $\psi_n^0$, with, say, $m$ and $n$ both odd, we get
after setting $\rho:=\sigma r$,
\begin{equation*}
\langle\psi_m^0 \vert |r| \vert \psi_n^0\rangle=-\frac{4\bar{a}%
_{\frac{m+1}{2}}a_{\frac{n+1}{2}}Ai^{\prime}\left(a_{\frac{m+1}{2}}\right)Ai^{\prime}\left(a_{%
\frac{n+1}{2}}\right)}{\sigma^2\left(a_{\frac{m+1}{2}}-a_{\frac{n+1}{%
2}}\right)}.
\end{equation*}
Using the expression for $a_n$, we see get the final result
\begin{equation*}
\langle\psi_m^0 \vert |r| \vert \psi_n^0\rangle=\frac{(-1)^{m+n+1}2}{\sigma\left(a_{%
\frac{m+1}{2}}-a_{\frac{n+1}{2}}\right)^2}.
\end{equation*}
For $m$ and $n$ both even, we have
\begin{equation*}
\langle\psi_m^0 \vert |r| \vert \psi_n^0\rangle=\frac{(-1)^{m+n}(a^{\prime}_{\frac{m+2}{2}%
}+a^{\prime}_{\frac{n+2}{2}})}{\sigma\left(a^{\prime}_{\frac{m+2}{2}}-a^{\prime}_{\frac{%
n+2}{2}}\right)^2\sqrt{a^{\prime}_{\frac{m+2}{2}}a^{\prime}_{\frac{n+2}{2}}}}.
\end{equation*}
For $m$ and $n$ both odd, we have
\begin{equation*}
\langle\psi_m^0 \vert r^2 \vert \psi_n^0\rangle=\frac{(-1)^{m+n+1}24}{\sigma^2\left(%
a_{\frac{n+1}{2}}-a_{\frac{m+1}{2}}\right)^4}.
\end{equation*}
For $m$ and $n$ both even, we have
\begin{equation*}
\langle\psi_m^0 \vert r^2 \vert \psi_n^0\rangle=\frac{(-1)^{m+n}12\left(a^{\prime}_{\frac{m+2%
}{2}}+a^{\prime}_{\frac{n+2}{2}}\right)}{\sigma^2\left(a^{\prime}_{\frac{n+2}{2}%
}-a^{\prime}_{\frac{m+2}{2}}\right)^4\sqrt{a^{\prime}_{\frac{m+2}{2}}a^{\prime}_{\frac{%
n+2}{2}}}}.
\end{equation*}

\section*{Acknowledgement}

This work was supported by the Natural Sciences \& Engineering Research
Council of Canada.

\end{document}